# LETTER

# A dusty, normal galaxy in the epoch of reionization

Darach Watson[1], Lise Christensen[1], Kirsten Kraiberg Knudsen[2], Johan Richard[3], Anna Gallazzi[4,1], and Michał Jerzy Michałowski[5]

Candidates for the modest galaxies that formed most of the stars in the early universe, at redshifts $z > 7$, have been found in large numbers with extremely deep restframe-UV imaging[1]. But it has proved difficult for existing spectrographs to characterise them in the UV[2,3,4]. The detailed properties of these galaxies could be measured from dust and cool gas emission at far-infrared wavelengths if the galaxies have become sufficiently enriched in dust and metals. So far, however, the most distant UV-selected galaxy detected in dust emission is only at $z = 3.2$[5], and recent results have cast doubt on whether dust and molecules can be found in typical galaxies at this early epoch[6,7,8]. Here we report thermal dust emission from an archetypal early universe star-forming galaxy, A1689-zD1. We detect its stellar continuum in spectroscopy and determine its redshift to be $z = 7.5±0.2$ from a spectroscopic detection of the Lyα break. A1689-zD1 is representative of the star-forming population during reionisation[9], with a total star-formation rate of about 12 $M_\odot$ yr$^{-1}$. The galaxy is highly evolved: it has a large stellar mass, and is heavily enriched in dust, with a dust-to-gas ratio close to that of the Milky Way. Dusty, evolved galaxies are thus present among the fainter star-forming population at $z > 7$, in spite of the very short time since they first appeared.

As part of a programme to investigate galaxies at $z > 7$ with the X-shooter spectrograph on the Very Large Telescope, we observed the candidate high-redshift galaxy, A1689-zD1, behind the lensing galaxy cluster, Abell 1689 (Fig. 1). The source was originally identified[10] as a candidate $z > 7$ system from deep imaging with the *Hubble* and *Spitzer* Space Telescopes. Suggested to be at $z = 7.6±0.4$ from photometry fitting, it is gravitationally magnified by a factor of 9.3 by the galaxy cluster[10], and though intrinsically faint, because of the gravitational amplification, is one of the brightest candidate $z > 7$ galaxies known. The X-shooter observations were carried out on several nights between March 2010 and March 2012 with a total time of 16 hours on target.

The galaxy continuum is detected and can be seen in the binned spectrum (Fig. 2). The Lyα cutoff is at 1035±24 nm and defines the redshift, $z = 7.5±0.2$. It is thus one of the most distant galaxies known to date to be confirmed via spectroscopy, and the only galaxy at $z > 7$ where the redshift is determined from spectroscopy of its stellar continuum. The spectral slope is blue; using a power-law fit, $F_\lambda \propto \lambda^{-\beta}$, $\beta = 2.0 \pm 0.1$. The flux break is sharp, and greater than a factor of ten in depth. In addition, no line emission is detected, ruling out a different redshift solution for the galaxy. Line emission is excluded to lensing-corrected depths of $3\times10^{-19}$ erg cm$^{-2}$ s$^{-1}$ (3σ) in the absence of sky emission lines, making this by far the deepest intrinsic spectrum published of a reionization era object, highlighting the difficulty of obtaining UV redshifts for objects at this epoch that are not strongly dominated by emission lines. The restframe equivalent width limits are < 4 Å for both Lyα and CIII] 1909Å. Our search space for Lyα is largely free of sky emission lines; they cover 16% of the range.

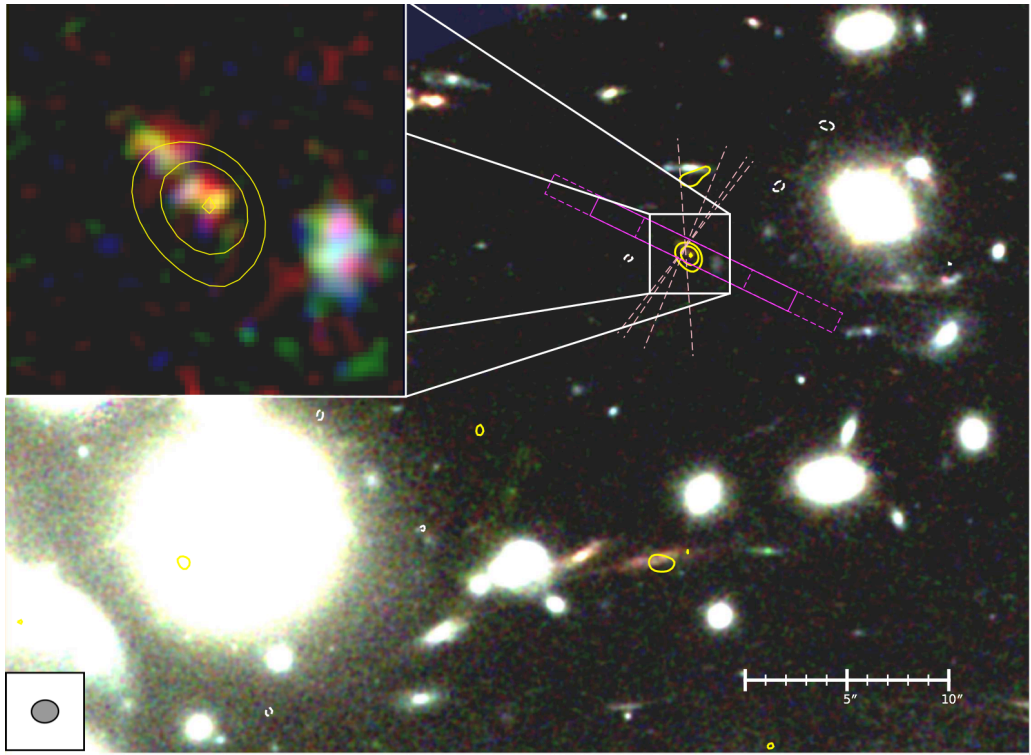

**Figure 1 | The gravitationally lensing galaxy cluster Abell 1689.** The colour image is composed with *Hubble* filters: F105W (blue), F125W (green), F160W (red). The zoomed box (4″×4″) shows A1689-zD1. Contours indicate FIR dust emission detected by ALMA at 3, 4, and 5σ local RMS (yellow, positive; white, negative). The ALMA beam (1.36″×1.15″) is shown, bottom left. Images and noise maps were primary-beam corrected before making the signal-to-noise ratio (SNR) maps. Slit positions for the first set of X-shooter spectroscopy are overlaid in magenta (dashed boxes indicate the dither), while the parallactic angle was used in the remaining observations (pink dashed lines).

[1]Dark Cosmology Centre, Niels Bohr Institute, University of Copenhagen, Juliane Maries Vej 30, København Ø, 2100, Denmark. [2]Department of Earth and Space Sciences, Chalmers University of Technology, Onsala Space Observatory, SE-439 92 Onsala, Sweden. [3]Centre de Recherche Astrophysique de Lyon, Observatoire de Lyon, Université Lyon 1, 9 Avenue Charles André, 69561 Saint Genis Laval Cedex, France. [4]Istituto Nazionale di Astrofisica–Osservatorio Astrofisico di Arcetri, Largo Enrico Fermi 5, 50125 Firenze, Italy. [5]The Scottish Universities Physics Alliance, Institute for Astronomy, University of Edinburgh, Royal Observatory, Edinburgh, EH9 3HJ, UK

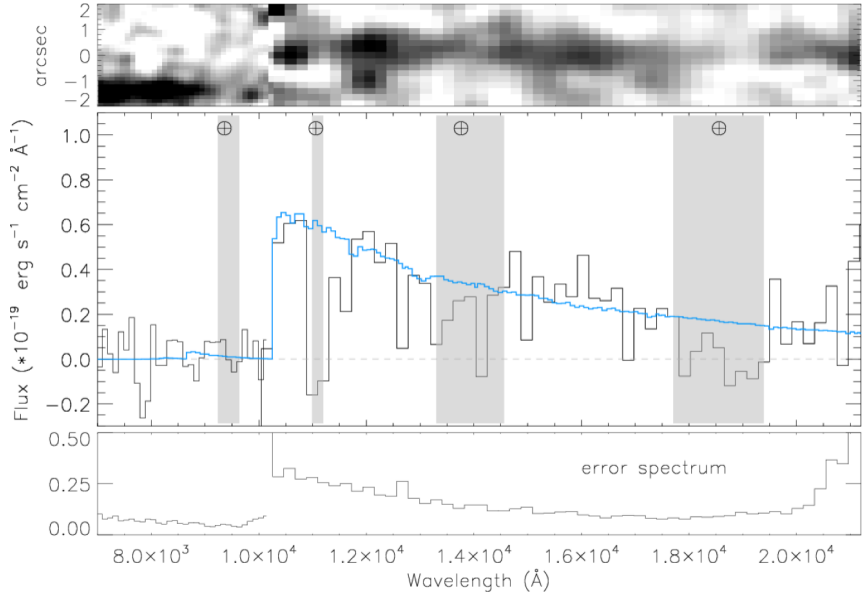

**Figure 2 | Spectrum of A1689-zD1.** The 1D (lower) and 2D (upper) binned spectra are shown, with the 68% confidence uncertainty on the 1D spectrum in the bottom panel. The redshift $z = 7.5$ is determined from the Lyα break at 1035 nm. Sky absorption (grey bands) and the best-fit SED (blue line) are shown. The Lyα break is close to the spectrograph's NIR/VIS arm split, however, the break is clearly detected in the NIR arm alone. A nearby galaxy ($z \sim 2$) visible in the bottom part of the 2D spectrum is detected in both the VIS and NIR arms.

Fits to the galaxy's spectral energy distribution (SED) yield a lensing-corrected stellar mass of $1.7 \times 10^9\,M_\odot$ (i.e. $\log(M_\star/M_\odot) = 9.23^{+0.15}_{-0.16}$, with a best-fitting stellar age of 80 Myr (i.e. a light-weighted age, $t$, of $\log(t/\mathrm{yr}) = 7.91^{+0.26}_{-0.24}$. The lensing-corrected UV luminosity is $\sim 1.8 \times 10^{10}\,L_\odot$, resulting in a star-formation rate estimate of $2.7 \pm 0.3\,M_\odot\,\mathrm{yr}^{-1}$ based on the UV emission and uncorrected for dust extinction, for a Chabrier initial mass function[11]. A1689-zD1 is thus a sub-$L^*$ galaxy, meaning that it is among the faint galaxies that dominate star-formation at this epoch[9].

Mosaic observations of the lensing cluster were obtained with the Atacama Large Millimetre Array (ALMA) in Cycles 0 and 1 with the receivers tuned to four 3.8 GHz frequency bands between 211 and 241 GHz. A1689-zD1 is located towards the northern edge of the mosaic and is detected at 5.0σ with an observed flux of 0.61 ± 0.12 mJy in the combined image and at 2.4–3.1σ significance in each of the three individual observations (Fig. 3). A1689-zD1 is located within the primary beam FWHM of one pointing and the sensitivity (RMS) around its position is 0.12 mJy/beam, 42% of the sensitivity of the deepest part of the mosaic. The source is the brightest in the mosaic area of 5 square arcminutes. It coincides with the UV position of A1689-zD1 and is 1.5″ away from the next nearest object in the *Hubble* image, which is not detected in the ALMA map. No line emission is convincingly detected in the ALMA spectral data, for which we cover about half of the X-shooter–allowed redshift space for the [C II] 158µm line (see Methods). The derived total infrared (TIR) luminosity corrected for lensing and for CMB effects[12] is $6.2 \times 10^{10}\,L_\odot$, corresponding to a SFR of about 9 $M_\odot\,\mathrm{yr}^{-1}$, with a dust mass in the galaxy of $4 \times 10^7\,M_\odot$, assuming a dust temperature of 35 K. The dust mass we determine is conservative and is unlikely to be lower than about $2 \times 10^7\,M_\odot$. It could in principle be larger by an order of magnitude, due to temperature and geometry effects (see Methods). Table 1 presents a comparison of the properties of A1689-zD1 with other $z > 6.5$ galaxies with deep mm observations. Most of the sources observed so far are more UV-luminous than A1689-zD1 and have upper limits somewhat above the dust mass found here.

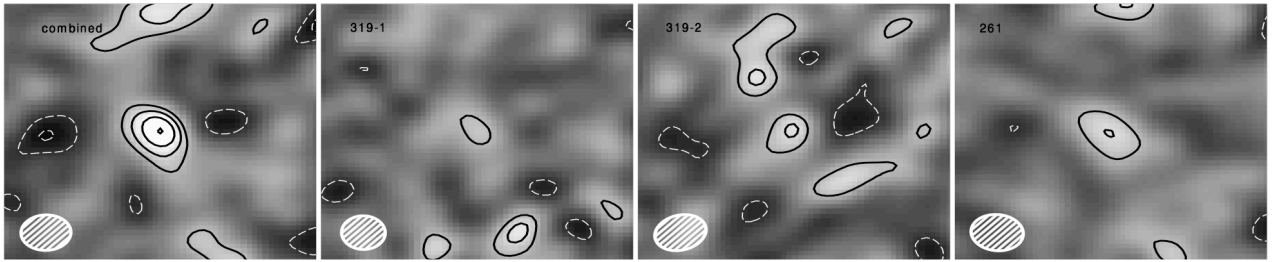

**Figure 3 | ALMA signal-to-noise ratio (SNR) maps of A1689-zD1.** Contours are SNR = 5, 4, 3, 2 (black, solid), –3, –2 (white, dashed). Images and noise maps were primary-beam corrected before making SNR maps. Beam sizes are shown, bottom left in each panel. Panels are 8″×8″. The panels show, left to right, the combined data, the two tunings of observation 2011.0.00319.S and observation 2012.1.00261.S. A1689-zD1 is detected, left to right, at 5.0σ, 2.4σ, 3.1σ, and 3.0σ. Natural weighting was used and the visibilities were tapered with a 1″ circular Gaussian kernel, resulting in beams of 1.36″×1.15″, 1.19″×1.09″, 1.43″×1.12″, 1.43″×1.17″ left to right.

**Table 1 | Comparison of A1689-zD1 to other high redshift star-forming galaxies.**

| Galaxy name | Redshift, $z$ | Stellar mass, $M_\star$ ($10^9 M_\odot$) | $SFR_{UV}$ ($M_\odot$ yr$^{-1}$) | $SFR_{Ly\alpha}$ ($M_\odot$ yr$^{-1}$) | $SFR_{IR}$ ($M_\odot$ yr$^{-1}$) | Dust mass, $M_D$ ($10^7 M_\odot$) |
|---|---|---|---|---|---|---|
| HFLS3 (ref. 21) | 6.34 | $50^{+100}_{-30}$ | $1.3 \pm 0.4$* | — | $1300_{-520}$† | $30_{-10}$† |
| HCM6A (ref. 22) | 6.56 | — | $9 \pm 2$ | 2 | <28 (ref. 26) | <10 (ref. 26) |
| Himiko (ref. 8) | 6.60 | $15 \pm 2$ | $30 \pm 2$ | $35 \pm 1$ | <8 | <4.72 (ref. 26) |
| A1703-zD1 (ref. 23) | 6.8 | 0.7–1.5 | $7.3 \pm 0.3$ | — | <16 (ref. 26) | <5.7 (ref. 26) |
| IOK-1 (ref. 24) | 6.96 | <40 | $23.9 \pm 1.4$ (ref. 27) | $10 \pm 2$ | <10 (ref. 28) | <6.4 (ref. 28) |
| z8-GND-5296 (ref. 2) | 7.51 | $1^{+0.2}_{-0.1}$ | $330^{+710}_{-10}$ | — | <127 (ref. 26) | <50 (ref. 26) |
| HG090423 (ref. 25) | 8.2 | <0.05 (ref. 29) | <0.38 (ref. 30) | — | <5 (ref. 29) | <2‡ |
| **A1689-zD1** | **7.5** | $1.7^{+0.7}_{-0.5}$ | $2.7 \pm 0.3$ | <0.7 | $9^{+4}_{-2}$ | $4^{+4}_{-2}$ |

The SFRs are derived from extinction-uncorrected ultraviolet emission, Ly$\alpha$ emission and far-infrared emission, respectively.
*Derived from the Hubble F160W photometry and corrected for lensing.
†95% lower bound only.
‡Assuming the same dust parameters assumed for A1689-zD1

Far-infrared emission requires the production of metals, whether in the solid phase as dust, or as ionised gas; to be detected, such galaxies must have enriched their interstellar media with metals and dust. While the metals are primarily produced and distributed via supernova explosions and so metal enrichment happens concurrently with massive star formation, the site of dust production is less certain. However, the mechanism must be very rapid[13] and these observations of A1689-zD1 place the strongest direct constraints so far on the rapidity of dust enrichment, occurring within only 500 Myr of the beginning of star-formation in the universe.

We can derive an approximate gas mass for this galaxy by inverting the Schmidt-Kennicutt law based on its size and star-formation rate[14] and we deduce a gas mass of $2^{+2}_{-1} \times 10^9 M_\odot$. This gives a dust-to-gas mass ratio of about $17 \times 10^{-3}$. And while the uncertainty on the gas and dust masses is large—approximately 0.5 dex, dominated by the scatter in the Schmidt-Kennicutt law and the unknown dust temperature, where the two values are linked through the SFR—the dust-to-gas mass ratio is nevertheless high for this redshift, between a half and a few times the Milky Way value[15]. For a constant dust-to-metals ratio[16], this also suggests a high metallicity with a similar uncertainty. Such a high metallicity is consistent with the fundamental mass-metallicity-SFR relation for star-forming galaxies[17], from which we determine an expected elemental abundance ~50% of the solar value. Instead of a young, dust-poor galaxy, these measurements suggest an evolved system. Finally, the deep upper limit on the C III] 1909Å emission line, of <4 Å restframe equivalent width is unusual for line-emitting galaxies at these redshifts[18]. Based on galaxies at lower redshift, we might expect an equivalent width as high as 30 Å for this star-formation rate for a young galaxy[19]. And while the lack of Ly$\alpha$ emission in this galaxy could be explained by IGM absorption, the absence of C III] emission cannot, and is consistent with a more evolved galaxy. The surface area we are observing in UV emission is only approximately 1.5 kpc$^2$ (lensing-corrected size), thus the star-formation rate per unit area is high, approximately 8 $M_\odot$ yr$^{-1}$kpc$^{-2}$, comparable to vigorous starburst galaxies with similar SFRs[14], but an order of magnitude or more lower per unit surface area than the extreme starbursts found in high redshift submm surveys or QSOs[20].

The comparison of the gas to stellar mass in this galaxy shows about $55 \pm 25$% of the baryonic matter in the form of gas, indicating that the galaxy has already formed much of its stars and metals. Taken together, these lines of evidence point to a picture of A1689-zD1 consistently forming stars at a moderate rate since $z \sim 9$, or possibly having passed through its extreme starburst very rapidly and now in a declining phase of star-formation.

It has been suggested that the decreasing metal contents of high redshift galaxies will make them challenging to detect at far-infrared wavelengths[6,7]. A1689-zD1 is at $z = 7.5$ and though it is magnified by a factor of 9.3, it was detected in only brief observations with ALMA. This promises a reasonable detection rate for $L^*$ galaxies in unlensed fields at these redshifts for the full ALMA array, in contrast to the gloomy outlook painted by observations of very low metallicity systems at high redshift[7,8]. The precise identification and detailed characterisation of the early universe star-forming population should therefore be possible in the far-infrared in the near future and should not be restricted to rare hyperluminous infrared galaxies.

**Acknowledgements** The Dark Cosmology Centre is funded by the Danish National Research Foundation. L. C. is supported by the EU under a Marie Curie Intra-European Fellowship, contract PIEF-GA-2010-274117. K. K. acknowledges support from the Swedish Research Council and the Knut and Alice Wallenberg Foundation. J. R. acknowledges support from the ERC starting grant, CALENDS, and the Career Integration Grant 294074. AG acknowledges support from the European Union Seventh Framework Programme (FP7/2007-2013) under grant agreement n. 267251 ("AstroFIt"). MJM acknowledges the support of the Science and Technology Facilities Council. ALMA is a partnership of ESO (representing its member states), NSF (USA) and NINS (Japan), together with NRC (Canada) and NSC and ASIAA (Taiwan), in cooperation with the Republic of Chile. The Joint ALMA Observatory is operated by ESO, AUI/NRAO and NAOJ. We thank Lukas Lindroos, Jens Hjorth, Johan Fynbo, Anja C. Andersen, and Rychard Bouwens for useful discussions, Marceau Limousin for providing a lensing map of the cluster, and the Nordic ALMA Regional Center Node for assistance.

**Author Contributions:** D. W. conceived the study, was Principal Investigator of the X-shooter programme, produced Fig. 1 and Extended Data Figs. 1 and 4–7 and wrote the main text. L. C. reduced and analysed the X-shooter spectrum, did the HyperZ analysis and produced Fig. 2 and Extended Data Fig. 2. K. K reduced and analysed the ALMA data and produced Fig. 3 and Extended Data Fig. 3. J. R. was Principal Investigator of the ALMA programmes and reduced and analysed the *Hubble* data. A. G. modelled the UV SED and determined the galaxy stellar age. M. J. M. modelled to the full UV-FIR SED and produced Table 1. All authors contributed to the Methods and all authors discussed the results and commented on the manuscript.

**Author Information** This paper makes use of the following ALMA data: ADS/JAO.ALMA 2011.0.00319.S and 2012.1.00261.S. Reprints and permissions information is available at www.nature.com/reprints. The authors declare no competing financial interests. Correspondence and requests for materials should be addressed to D. W. (darach@dark-cosmology.dk)

## METHODS

All uncertainties quoted are 68% confidence unless stated otherwise.

**Optical spectroscopy.** The X-shooter spectrograph on the Very Large Telescope was used to observe the source A1689-zD1 on the nights 21 March and 18 April 2010, as well as shorter exposures on 27 March, 25 April, 1, 2, 3 May 2011, and 17 and 24 March 2012. The observations used a standard dither pattern. Approximately half the exposure time (the 2010 observations) was performed with a fixed slit position angle, while the rest used the parallactic angle. The slit setups are indicated in Fig. 1. The data were reduced with the ESO X-shooter pipeline version 2.2. The standard calibration routines were used, with the different nod positions employed to subtract the background sky emission. Spectrophotometric standard stars observed on the same night as the galaxy were used for flux calibration. The target was acquired using an offset from a bright nearby star, calculated from the *Hubble* imaging. Offsets are executed with an accuracy of <0.1″, i.e. the location of the galaxy in the slit is known a priori to better than a pixel. The slit-centre position and offsets along the slit were used to shift and co-add the data into a final combined 2D spectrum.

**Spectroscopic redshift.** The spectroscopy rules out a low redshift solution. We detect no emission lines in the spectrum (see below), but the stellar continuum of the galaxy is detected, allowing us to measure the blue UV continuum and a sharp break, which cannot be reproduced with dust extinction. Other breaks are excluded by the sharpness and depth of the break in spectroscopy, the blue slope of the UV continuum, and the deep upper limits in the VIS arm of the spectrograph. These data are independent of the original discovery data for the dropout, removing any Eddington-bias–like effect and improving the reliability of the redshift determination.

The SNR per spectral pixel is low. We therefore bin the X-shooter spectra by a large factor in wavelength (see ref. 31 for details). We fit a template spectrum of a ~100 Myr galaxy to estimate the redshift. This template was shifted to a range of redshifts and corrected for the transmission in the intergalactic medium[32]. The best-fit redshift was found from $\chi^2$ minimization. We tested the fit with a variety of bin-sizes: 150, 180, 200, 250 and 400 pixels in the NIR. The VIS data bin-sizes are 1.5 times larger. We extracted the 1D spectrum for each binning and performed 1000 realisations of the model where we added Gaussian noise corresponding to the uncertainty in each binned pixel. We calculated the average of the five binning sizes and determined the average 68% confidence level to be $z = 7.5\pm0.2$.

To be confident of the location of the spectral break, we also adopt a standard methodology for detecting steps in one-dimensional data: searching for a change in slope in the unbinned cumulative sum (Extended Data Fig. 1). The break at ~1035 nm is clearly observed only in the NIR arm data. We simulated the data 500 times, using the error spectrum as the standard deviation for the Gaussian random realisations, and following the same cumulative sum method and data-cleaning as for the real data, we derived the break position and its uncertainty. This is very close to the analysis of the binned data: $z = 7.5\pm0.2$. The fit results and error are independent of the VIS arm data and depend only on the NIR spectrum.

A galaxy at $z \sim 2$ is reported[10] offset by 1.5″ from the lensed galaxy. In Extended Data Fig. 2 we show its undiluted spectrum from the 2010 data alone. Its spectrum extends across the NIR/VIS arm boundary and does not show the sharp break seen in A1689-zD1. In Fig. 2 (main text), the spectrum of this nearby galaxy is diluted because the spectra are observed at different slit orientations, some of which did not cover the companion (see Fig. 1).

Finally, we fit the full set of photometric data (all *Hubble* and *Spitzer* bands) with the New-HyperZ code[33], version 12.2, to determine a photometric redshift, for completeness. Only one redshift solution is allowed, at $z \sim 7.4$ (Extended Data Fig. 3), consistent with the spectroscopic determination.

**Emission line flux detection limits.** No emission line is found from visual inspection of the spectrum. Of the strong emission lines, at $z = 7.5$ only Lyα could in principle be detected (He II 1640 Å would be too faint). We estimated the limits for the detection of any emission line by adding artificial emission lines at random within the redshift range determined from the spectral break redshift calculation. These lines were simulated with FWHMs in the range 50–200 km s$^{-1}$ and were added to the unbinned 2D spectrum. The 1D spectra were extracted and binned by 3–5 pixels. A Gaussian function was fit to the data and the SNR of the artificial line flux derived.

We place a 3σ upper limit to the Lyα emission line of $1.8\times10^{-17}$ erg cm$^{-2}$ s$^{-1}$ at 10030 Å near the sky lines and a factor of seven lower away from the sky lines (not corrected for lensing magnification). This limit becomes tighter from the blue to the red, i.e. at 10330 Å and 10573 Å the limits near the sky lines are $0.9\times10^{-17}$ erg cm$^{-2}$ s$^{-1}$ and $0.5\times10^{-17}$ erg cm$^{-2}$ s$^{-1}$, respectively.

To determine the effective Lyα escape fraction, we first convert this limit to a nominal SFR by assuming a factor 8.7 flux ratio between Lyα and Hα, and a standard conversion between Hα and the star-formation rate[34], but assuming a Chabrier IMF[11], as used throughout the paper. This corresponds to a detection limit for the nominal Lyα SFR $<0.7\,M_\odot$ yr$^{-1}$ (corrected for lensing). Between the sky lines, the upper limit is $<0.09\,M_\odot$ yr$^{-1}$. This implies an effective Lyα escape fraction of either <6% or <0.8% by comparison to the total SFR of $12\,M_\odot$ yr$^{-1}$. The rest-frame Lyα EW is <27 Å, or <4 Å between the sky lines. We cannot distinguish between Lyα absorbed by the galaxy ISM, circumgalactic medium or IGM. However, the dustiness of the galaxy indicates a substantial fraction of the Lyα may be absorbed in the host.

Similarly, we place a 3σ upper limit of $2\times10^{-18}$ erg cm$^{-2}$ s$^{-1}$ for the lensing uncorrected flux of the CIII] 1909Å emission line, which lies in a region of the spectrum relatively free of atmospheric effects, corresponding to a restframe equivalent width of < 4 Å.

**The ALMA observations and data reduction.** We obtained ALMA mosaic observations for A1689 in Cycle 0 and Cycle 1 as part of the projects 2011.0.00319.S and 2012.1.00261.S. The receivers were tuned to cover 211.06–214.94 GHz, 221.46–225.34 GHz, 227.06–230.94 GHz, 237.46–241.34 GHz; the 2012.1.00261.S data used here cover the first and third frequency setup. The correlator was used in the frequency domain mode with a bandwidth of 1875 MHz in each spectral window. The projected baselines range between 12m and 450m. The quasar 3C 279 was used for bandpass and phase calibration. The distance from 3C 279 to A1689 is 5.9°. Mosaicked images were primary beam corrected and the weights of the individual fields were taken into account. Flux calibration was done using Mars and Titan.

Data reduction used the Common Astronomy Software Application[35] versions 3.4 and 4.1 for the Cycle 0 and 1 data respectively. For the Cycle 0 data, additional careful reduction of the observatory-provided preliminary-reduced data was required, including flagging of noisy data as well as producing a frequency dependent model of 3C 279 to account for the spectral index of the continuum emission through bootstrapping from the flux calibrators. Our results are, however, not more than 1σ different from the preliminary-reduced data.

The data were combined and imaged using CASA 4.2; when imaging, the calibrated visibilities were naturally weighted (resulting beam-size 0.8″×1″) and tapered using a 2D Gaussian with 1″×1″ to give greater weight to shorter baselines, resulting in a beam-size of 1.36″×1.15″. The full mosaic will be presented in a forthcoming paper (Knudsen et al., in preparation). The SNR images are shown in Fig. 3. The flux image for A1689-zD1 is shown in Extended Data Fig. 4. The data reduction scripts used for the ALMA data can be obtained at: http://dark.nbi.ku.dk/research/archive.

The most conservative estimate of the astrometric uncertainty is half the beam dimensions, i.e. 0.35″×0.5″, dominated by residual atmospheric phase effects and the signal-to-noise ratio. Together with four other ALMA detected sources, we found an average offset of 0.4–0.45″ between the ALMA centroids and the *Hubble* centroids. Part of these offsets is likely contributed by intrinsically different optical and far-infrared morphologies. No significant flux is detected from the galaxy 1.5″ from A1689-zD1.

No spectral line is detected toward A1689-zD1 in the ALMA data. Our data cover [C II] 158μm emission at: $z = 6.87$–7.00, 7.23–7.37, 7.43–7.58, 7.84–8.00. This covers ~50% of the 1σ and 2σ ranges allowed by X-shooter. The typical observed luminosity ratio $L_{[C II]}/L_{FIR}$ varies between 0.001 and 0.008[36]. Over the redshift ranges above, we exclude [C II] 158μm emission at the high end of this ratio (>0.0024, 5σ for a linewidth of 100 km s$^{-1}$).

**Dust mass and temperature.** We derive the FIR luminosity (42–122 μm restframe) and dust mass using a single temperature modified blackbody and assuming a typical value for the dust mass absorption coefficient of $\kappa_\nu = 0.067 \times \left[(1+z)\frac{\nu}{250}\right]^{\beta_{IR}}$ m$^2$ kg$^{-1}$ at the observed frequency $\nu = 226$ GHz, and $\beta_{IR} = 1.92$ (as observed in HFLS3[37]). Different values for the dust mass absorption coefficient could result in dust masses ±0.18 dex[38]. We explore the effect of using more complex full SED models in a section below.

With only one FIR flux point it was necessary to assume a value of the dust temperature to recover the dust mass. We assumed 35K. A lower temperature would increase the inferred dust mass, making the galaxy richer in dust than we infer. The galaxy may not be optically thin to FIR radiation; we assume the wavelength of optical depth unity is $\lambda_0 = 200$ μm. We apply corrections due to CMB heating and the error induced by the large CMB background subtraction[12]. We show the SFR and dust mass inferred using our fiducial modified blackbody model for various assumed dust temperatures, $\beta_{IR}$ and $\lambda_0$ values in Extended Data Fig. 5.

The constraints induced by the observed correlation between the UV spectral slope, $\beta_{UV}$, and the ratio of observed infrared to UV luminosity, IRX[39], with the flux uncertainties and $\beta_{UV}$ slope uncertainties included, are also shown in Extended Data Fig. 5. The fiducial dust mass is $\log(M_D/M_\odot) = 7.6$, with a 2σ lower bound of 7.30, and an upper bound that is not strongly constrained, but where $\log(M_D/M_\odot) = 7.9$ implies a dust-to-gas ratio approximately five times that of the Milky Way (see below), and is therefore disfavoured. It has been shown recently that galaxies at high redshift may be preferentially more dusty[40]. This higher IRX, by 0.3 dex, is quite consistent with A1689-zD1.

**Star-formation rate.** We measure the UV flux ($F_\nu$) from the *Hubble* F160W photometric data point. The UV luminosity is calculated for the best-fit redshift of 7.5 using our standard cosmological parameters[41]. No dust correction is applied in calculating the UV SFR. The SFR is derived from the UV and total IR (TIR, 3–1100 μm) luminosities[33], $L_{UV}$ and $L_{TIR}$. For the UV luminosity we find $L_{UV} = 1.8 \pm 0.2 \times 10^{10} L_\odot$, corresponding to a SFR = 2.7 ± 0.3 $M_\odot$ yr$^{-1}$. For the TIR luminosity we obtain $L_{TIR} = 6.2 \pm 0.8 \times 10^{10} L_\odot$, which is the flux uncertainty only, using our fiducial model, but where the true uncertainty is constrained primarily by the desire to remain consistent within the scatter with the IRX-$\beta_{UV}$ relation. These values are consistent with those derived from modelling the SED (see below) and result in a SFR$_{TIR} = 9 \pm 2$ $M_\odot$ yr$^{-1}$, using recent calibrations based on the 3–1100 μm IR luminosity[14]. A SFR$_{TIR}$ much below 7 $M_\odot$ yr$^{-1}$ is excluded in this model. A significantly lower dust temperature could result in a higher SFR$_{TIR}$, but would imply an excessive dust mass. From the analysis in Extended Data Fig 5, our best estimate is SFR$_{TIR} = 9^{+4}_{-2}$ $M_\odot$ yr$^{-1}$.

The sum of the UV and infrared SFRs results in a total SFR of $12^{+4}_{-2}$ $M_\odot$ yr$^{-1}$, for this IMF (statistical uncertainty only), with a likely additional contribution to the luminosity and hence the SFR from the mid-IR emission that is not well-modelled by a single temperature modified blackbody (see SED modelling below). Therefore, we take this as a conservative estimate of the SFR and note that it may be somewhat higher.

**Gas mass and metallicity.** We invert the Schmidt-Kennicutt law relating the surface densities of SFR and gas mass[14] to obtain constraints on the gas mass. The dominant uncertainty on this conversion comes from the uncertainty in the slope of the S-K law, which is a ±0.4 dex scatter—the dependence on the surface area of the galaxy is very small (0.04 dex) since it appears in both the SFR and gas mass terms. We here use the relation as originally derived, which reproduces even starburst galaxies quite well, assuming a constant CO to gas conversion, X(CO)[14]. Low metallicity galaxies may have lower gas mass for a given SFR[14], which would only increase the dust-to-gas ratio derived here. The derived gas mass is inversely dependent on the inferred lensing magnification. Since A1689-zD1 is not very close to the critical lines, its lensing amplification is likely fairly accurate. A different analysis of the data for the cluster gives a similar magnification of 8.6±0.5 for this source (M. Limousin, private communication), confirming that the lens model contributes only a small additional uncertainty, 0.025 dex, for the calculations presented here. The overall uncertainty in the gas mass summed in quadrature is therefore 0.45 dex, including the scatter in the S-K relation and the uncertainty on the SFR.

The dust-to-gas ratio we derive for our fiducial model is 0.017, with a total uncertainty of 0.5 dex under the assumptions outlined above for the derivation of the dust and gas masses. Both values depend on the SFR and this is accounted for in the uncertainty on the ratio. The scatter in the S-K relation and the dust mass systematics dominate the uncertainty.

**The age, mass and SED of A1689-zD1.** We construct the photometric SED of A1689-zD1 from *Hubble*'s ACS F775W and F850W, WFC3 F105W, F125W, F140W and F160W, and NIC2 F110W bands, and *Spitzer*'s IRAC 3.6 μm and 4.5 μm bands. Since its first discovery with *Hubble*, A1689-zD1 was observed on April 2010 with *Hubble*/WFC3 in the F105W, F125W, F140W and F160W bands, for 2.5 ks each (Proposal ID 11802, PI: Ford). The individual reduced frames were combined using the MultiDrizzle[42] software onto a common pixel scale of 0.05″ per pixel and the astrometry was matched to the ACS images. Isophotal magnitudes have been measured using SExtractor in double-image mode, taking the F140W image as a reference for the detection. The fluxes for the source from the newer *Hubble* imaging are; F105W: 25.49±0.18, F125W: 25.00±0.13, F140W: 24.64±0.05, F160W: 24.51±0.11. In addition, photometry from ref. 10 is used.

We build a large library of model galaxy SEDs by convolving Simple Stellar Population models[43] with randomly generated star formation histories. These are modelled with exponentially declining laws (parameterized by the formation redshift—bound to be younger than the Universe's age at $z = 7.5$—and the star formation timescale) with stochastic bursts of star formation (parameterized by their duration and fraction of mass formed). The models span a metallicity range between 20% and 2.5 times solar. Dust attenuation is applied to the stellar SEDs adopting a two-component model[44] and randomly generating the values for the total optical depth and the fraction of it contributed by the diffuse interstellar medium. Transmission in the intergalactic medium is finally applied[32].

Following the Bayesian approach developed and extensively applied to lower-redshift data[45,46] the observed SED is compared to every model SED in the library. This allows us to construct the full probability density function (PDF) of parameters of interest by weighting each model by its likelihood exp($-\chi^2/2$) and marginalizing over the nuisance parameters of the SFHs. For completeness we also derive the total (dust-corrected and averaged over the galaxy age) SFR, $9^{+5}_{-3}$ $M_\odot$ yr$^{-1}$, and the dust attenuation, $A_{1600} = 1.0 \pm 0.4$ mag, consistent with the values derived elsewhere in the paper. Because of the almost complete absorption of the flux below the Lyα by the

IGM, including the ACS upper limits or not does not change the fit. The best-fit to the UV-optical SED is shown in Extended Data Fig. 6.

Finally, strong emission lines are not included in the SED modelling. We consider possible contamination by the Balmer Hγ and Hδ lines to the 3.6 μm flux and the contamination by the Hβ and [OIII] lines to the 4.5 μm flux. Adopting a SFR of 12 $M_\odot$ yr$^{-1}$ and standard case B Balmer line ratios, we expect the 3.6 μm flux contribution to be at most 7%. The nebular line contribution to the 4.5 μm flux could amount to 8–14% (depending on the [OIII]/Hβ line ratio). By applying these corrections to the observed fluxes, our SED fit would yield a stellar mass 0.06 dex lower and a light-weighted mean age 0.04 dex younger than obtained above.

As a final check, we fit self-consistent full UV to far-infrared SED models using the MAGPHYS[47] and GRASIL[48,49,50] codes. From these fits we derived the following lensing-corrected parameters: SFR = $10^{+4}_{-2}$ $M_\odot$ yr$^{-1}$, $\log(M_\star/M_\odot) = 9.3^{+0.2}_{-0.1}$, $M_D = 3^{+3}_{-2} \times 10^7$ $M_\odot$ (MAGPHYS); SFR = $9^{+34}_{-2}$ $M_\odot$ yr$^{-1}$, $\log(M_\star/M_\odot) = 9.4^{+0.6}_{-0.2}$, $M_D = 7^{+5}_{-2} \times 10^7$ $M_\odot$ (GRASIL). Applying the same order of CMB correction to these fits as to the modified blackbody would increase each parameter by about 30% (i.e. +0.12 dex). The fits with these codes are shown in Extended Data Fig. 7.

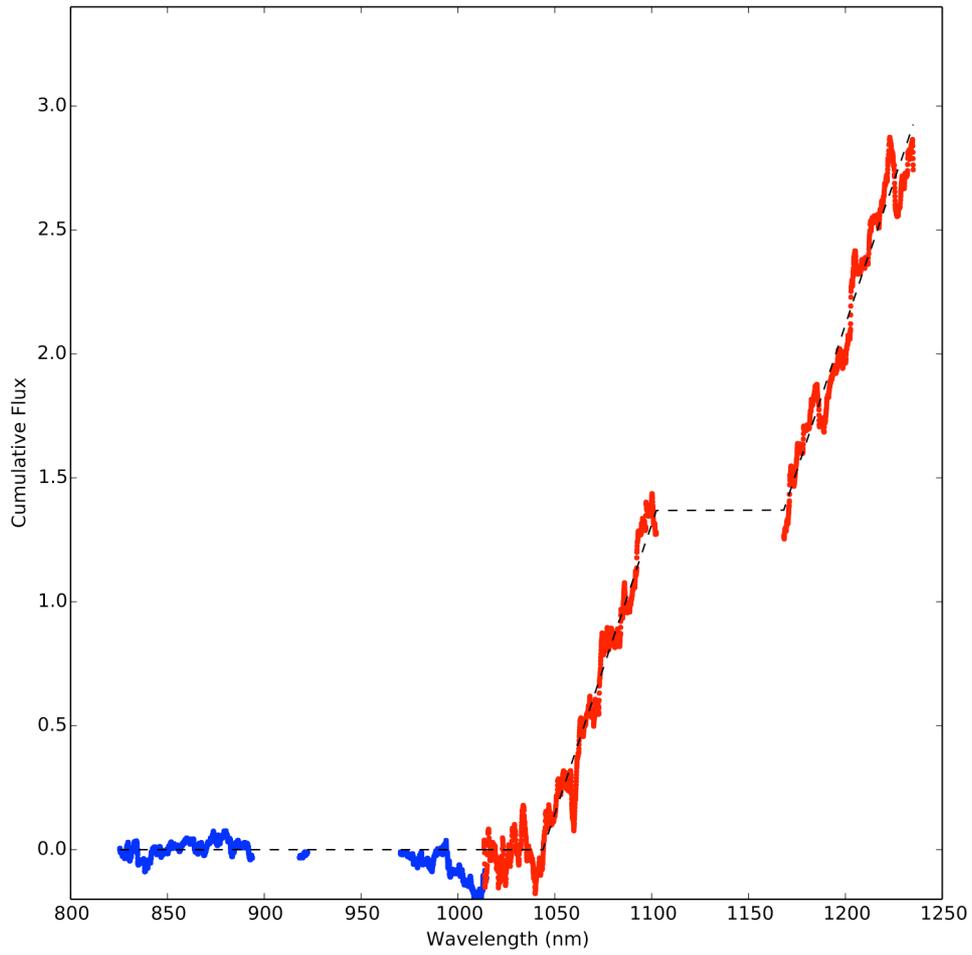

**Extended Data Figure 1 | Cumulative sum of the unbinned spectrum.** The VIS and NIR arms are plotted in blue and red respectively. The best fit step-function is plotted as a dashed line. The break in the spectrum is clearly detected with the NIR arm only. Gaps in the cumulative spectrum are due to removal of regions affected by strong sky absorption.

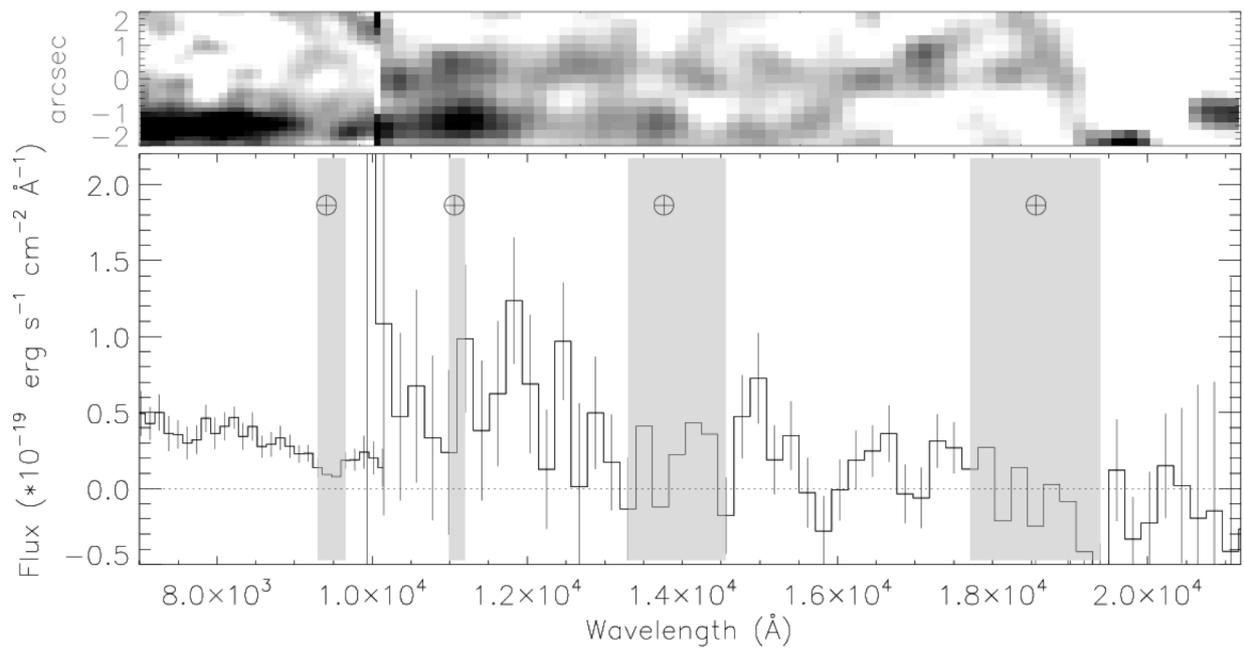

**Extended Data Figure 2 | Spectrum obtained only at position angle 64° East of North.** The slit consistently covered both the emission from the high redshift galaxy and the galaxy located 2″ below it. This spectrum uses approximately half the total exposure time. The upper panel shows the 2D rectified spectrum, the lower panel the 1D spectrum of the companion. Error bars are 68% confidence. The spectrum of the companion galaxy is recovered through the entire spectral range, also that covered by the transition from the VIS to the NIR data, and shows no indication of the sharp break seen in A1689-zD1.

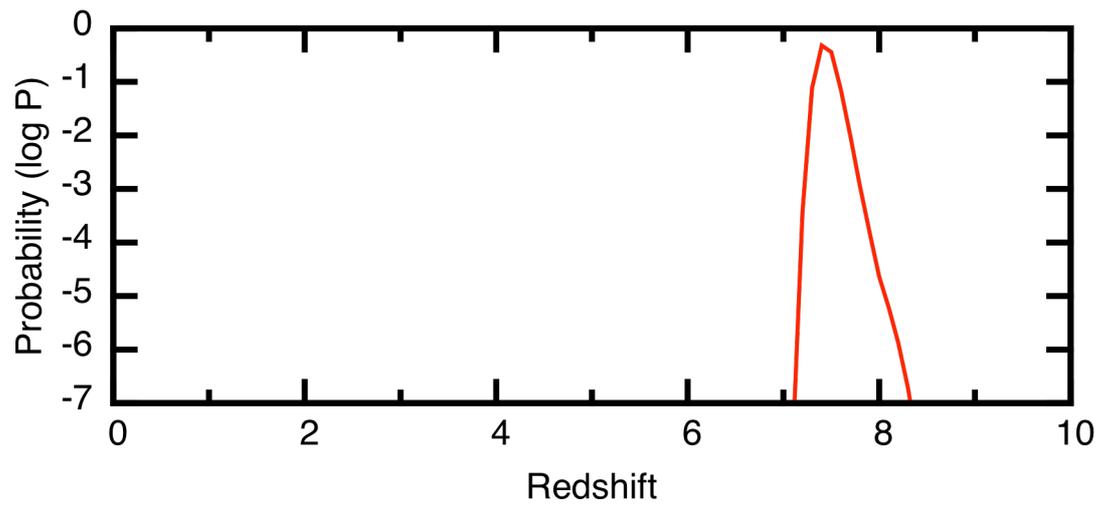

**Extended Data Figure 3 | Probability distribution as a function of redshift for galaxy template fits to the *Hubble* and *Spitzer*-IRAC photometry data.** The probabilities distribution is based on fitting galaxies using the New-HyperZ code.

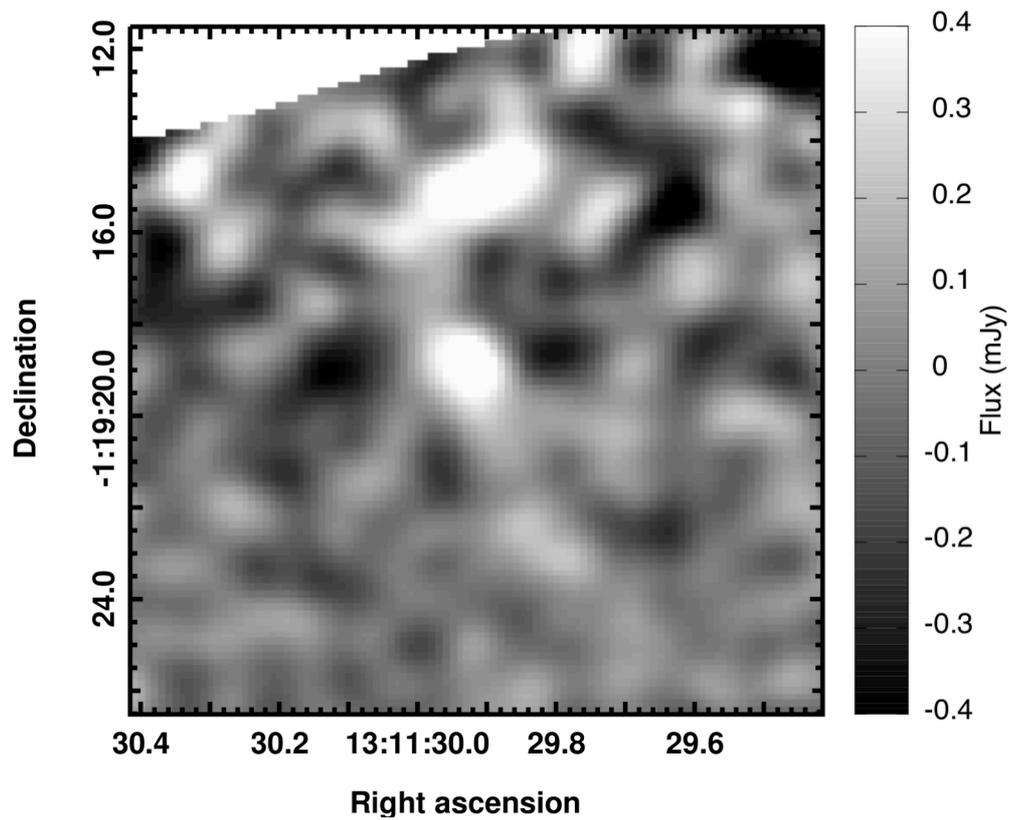

**Extended Data Figure 4 | The tapered ALMA flux image at 226 GHz, centered on A1689-zD1; the image is primary beam corrected.** The depth of the map at the location of zD1 is 0.12 mJy/beam (42% of the deepest part of the mosaic). The sensitivity decreases towards the edge of the mosaic due to the overlap of multiple pointings and primary beam correction. The structure north of zD1 is a probable detection of a different source in the field and will be presented in a forthcoming paper (Knudsen et al., in preparation).

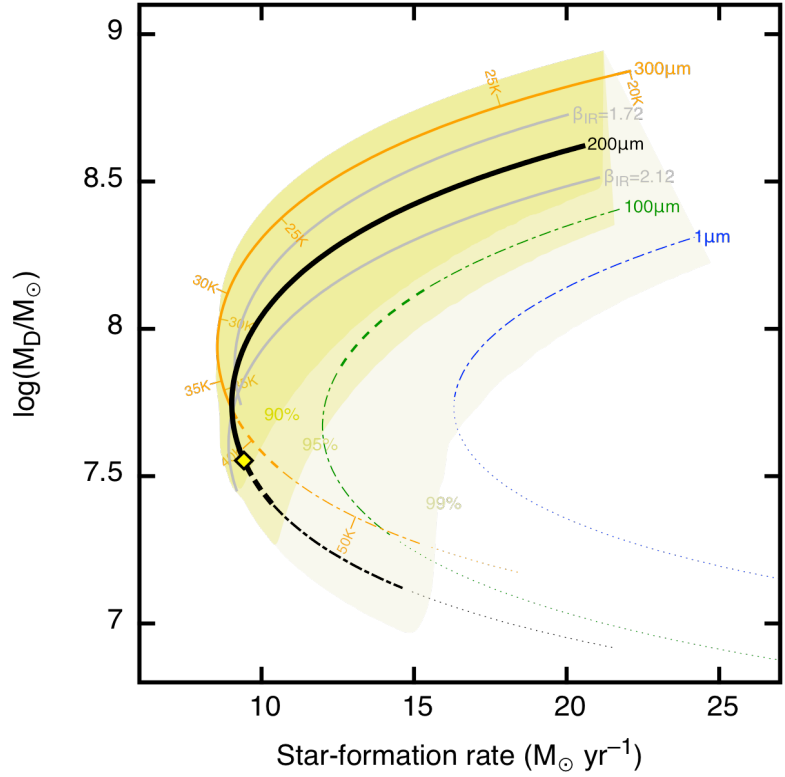

**Extended Data Figure 5 | Dust mass and SFR$_{TIR}$ from modified blackbody fits.** Tracks show how the parameters change with temperature, with different tracks for different opacity wavelengths ($\lambda_0$). Varying $\beta_{IR}$ is shown for $\lambda_0$=200µm (black and grey lines). Intrinsic (CMB-corrected) and measured temperatures are indicated for $\lambda_0$=300µm (orange line) on the concave and convex sides respectively. A diamond marks our fiducial model: measured T=35K, $\lambda_0$=200µm, $\beta_{IR}$=1.92. Solid-colour regions show <90%, <95%, and <99% confidence intervals due to the $\beta_{UV}$–IRX relation (including measurement uncertainties, $\beta_{IR}$=1.72–2.12, and $\lambda_0$<300µm), with solid, dashed, and dot-dashed lines indicating these intervals for the tracks. Dotted lines mark >99%.

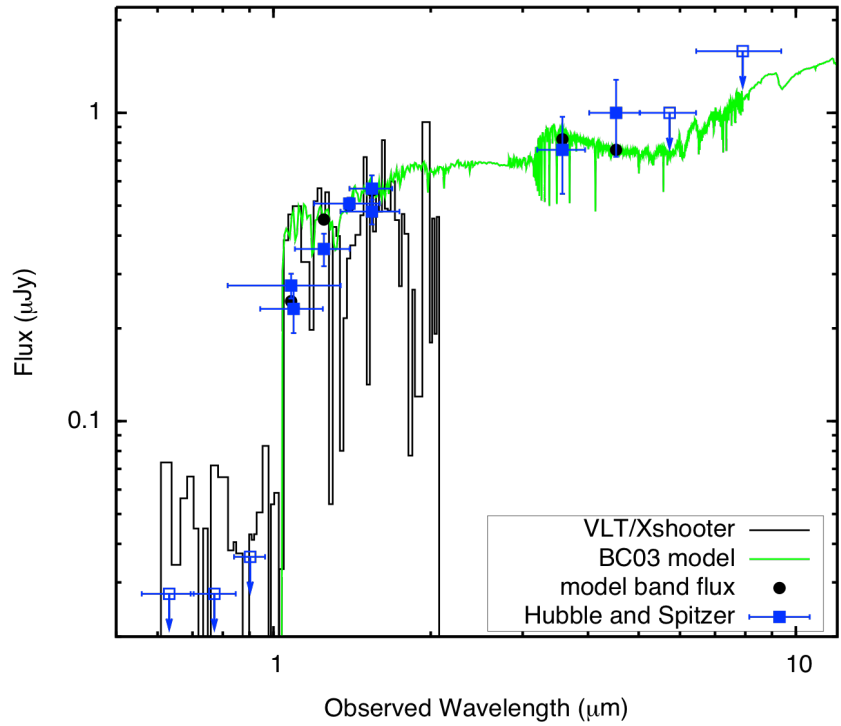

**Extended Data Figure 6 | UV-optical SED for A1689-zD1.** Bruzual & Charlot (2003) stellar synthesis models are fit to the photometric data (squares). Error bars are 68% confidence. The best-fit model is shown in green with resultant fluxes in the different bands shown as circles. The X-shooter spectrum is also plotted (solid histogram) for comparison.

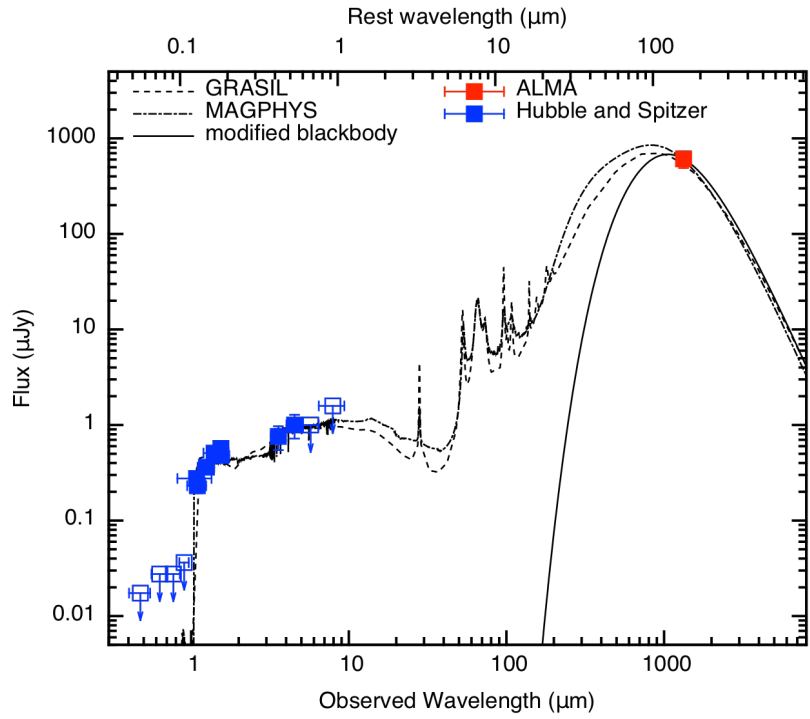

**Extended Data Figure 7 | SED of A1689-zD1.** Full, self-consistent UV-to-FIR models are fit to the data using the GRASIL (dashed line) and MAGPHYS (dot-dashed line) codes. The values derived from these models fit to the photometric data (squares) are largely consistent with those derived from the modified blackbody (solid line) and UV-optical only fit, though with an additional contribution from the restframe mid-IR flux. A CMB correction has not been applied here. Error bars are 68% confidence. Upper limits are 68% confidence except for the 8.0μm band for which the upper limit is 95% confidence.